\documentclass{jfm}
\pdfoutput=1
\usepackage{enumerate}
\usepackage{color}
\usepackage{graphicx}
\usepackage{natbib}
\usepackage{bm}
\usepackage{amsmath}
\usepackage[cal=rsfso,calscaled=.96]{mathalfa}

\usepackage[breaklinks]{hyperref}

\DeclareMathAlphabet{\mathpzc}{OT1}{pzc}{m}{it}
\DeclareMathAlphabet{\mathcalligra}{T1}{calligra}{m}{n}
\ifCUPmtlplainloaded \else
  \checkfont{eurm10}
  \iffontfound
    \IfFileExists{upmath.sty}
      {\typeout{^^JFound AMS Euler Roman fonts on the system,
                   using the 'upmath' package.^^J}%
       \usepackage{upmath}}
      {\typeout{^^JFound AMS Euler Roman fonts on the system, but you
                   dont seem to have the}%
       \typeout{'upmath' package installed. JFM.cls can take advantage
                 of these fonts,^^Jif you use 'upmath' package.^^J}%
      }
  \else
  \fi
\fi

\ifCUPmtlplainloaded \else
  \checkfont{msam10}
  \iffontfound
    \IfFileExists{amssymb.sty}
      {\typeout{^^JFound AMS Symbol fonts on the system, using the
                'amssymb' package.^^J}%
       \usepackage{amssymb}%

      }{}
  \fi
\fi

\ifCUPmtlplainloaded \else
  \IfFileExists{amsbsy.sty}
    {\typeout{^^JFound the 'amsbsy' package on the system, using it.^^J}%
     \usepackage{amsbsy}}
    {\providecommand\boldsymbol[1]{\mbox{\boldmath $##1$}}}
\fi
\newcommand{\aap}{Astron. Astrophys.}
\newcommand{\mnras}{Mon. Not. Roy. astron. Soc.}
\newcommand{\apjl}{Astrophys. J. Lett.}
\newcommand{\apj}{Astrophys. J.}
\newcommand{\apjs}{Astrophys. J. Suppl.}
\newcommand{\pre}{Phys. Rev. E}
\newcommand{\prl}{Phys. Rev. Lett.}
\newcommand{\jfm}{J. Fluid Mech.}

\providecommand\bnabla{\boldsymbol{\nabla}}
\providecommand\bcdot{\boldsymbol{\cdot}}

\newcommand\Rey{\mbox{\textit{Re}}}  

\newsavebox{\astrutbox}
\sbox{\astrutbox}{\rule[-5pt]{0pt}{20pt}}

\title[Scaling in supersonic turbulence]
{Energy cascade and scaling in supersonic isothermal turbulence}

\author[A. G. Kritsuk, R. Wagner and M. L. Norman]%
{Alexei G. Kritsuk$^1$%
  \thanks{Email address for correspondence: akritsuk@ucsd.edu},\ns
Rick Wagner$^2$
and Michael L. Norman$^{1,2}$}

\affiliation{\small $^1$Department of Physics and Center for Astrophysics and Space
Sciences, University of California, San Diego, MC 0424, 9500 Gilman
Drive, La Jolla, CA 92093-0424, USA\\[\affilskip]
$^2$San Diego Supercomputer Center, University of California,
San Diego, MC 0505, 10100 Hopkins Drive, La Jolla, CA 92093-0505, USA
}

\pubyear{2013}
\volume{7XX}
\pagerange{NNN--MMM}
\date{\today}

\begin{document}

\maketitle

\begin{abstract}
Supersonic turbulence plays an important role in a number of extreme astrophysical and terrestrial environments, yet its understanding
remains rudimentary. We use data from a three-dimensional simulation of supersonic isothermal turbulence
to reconstruct an exact fourth-order relation derived analytically from the Navier--Stokes equations
(Galtier and Banerjee, {\em Phys. Rev.  Lett.}, vol. 107, 2011, p. 134501). Our analysis supports a Kolmogorov-like inertial
energy cascade in supersonic turbulence previously discussed on a phenomenological level. We show that two
compressible analogues of the four-fifths law exist describing fifth- and fourth-order correlations, but only the fourth-order
relation remains `universal' in a wide range of Mach numbers from incompressible to highly compressible regimes.
A new approximate relation valid in the strongly supersonic regime is derived and verified. We also briefly discuss the origin
of bottleneck bumps in simulations of compressible turbulence.
\end{abstract}

\begin{keywords}
{\small compressible turbulence, turbulence simulation, turbulence theory, turbulent flows}
\end{keywords}
\section{Introduction}
Supersonic turbulence is believed to play a key role in a wide range of extreme astrophysical and terrestrial environments; for example, 
regulating star formation in molecular clouds \citep{hennebelle.12}, feeding supermassive black holes \citep{hobbs...11},  
creating clumpy structure in hot winds from Wolf-Rayet stars \citep{moffat.94}, 
controlling air entrainment in high-pressure volcanic eruptions \citep{ogden..08}, and affecting fuel mixing and combustion efficiency
in scramjets \citep{ingenito.10}. 

Compared to incompressible turbulence, highly
compressible turbulent flows are more complex due to nonlinear coupling of the velocity, density and pressure 
fields. Shock waves and vortex sheets change the topology of intermittent dissipative
structures in supersonic turbulence \citep{pan..09}. A `universal' scaling of the mass-weighted velocity $\bm v\equiv\rho^{1/3}\bm u$ 
was demonstrated in numerical experiments \citep{K07a,K07b} and independently verified in a number of numerical studies
\citep{kowal.07,schmidt..08,federrath....10,price.10,schwarz...10},
suggesting, by dimensional arguments, the presence of an inertial cascade. More recently, analytical scaling relations 
for compressible turbulence were derived and analyzed \citep{falkovich..10,galtier.11,wagner...12,banerjee.13} and the
existence of an intermediate scaling range dominated by inertial dynamics was demonstrated rigorously based on coarse-graining
\citep{aluie11,aluie..12,aluie13}. This contribution reports on the verification of the new relation presented in \citet{galtier.11} with data from a
Mach 6 simulation \citep{K07a} and on the phenomenology that follows from these results.

\section{A fourth-order relation}
Consider a system of Navier--Stokes equations for an isothermal compressible fluid in three dimensions
\begin{eqnarray}
\partial_t\rho+{\bnabla}\bcdot(\rho {\bm u}) & = & 0,\label{mass}\\
\partial_t(\rho {\bm u})+{\bnabla}\bcdot(\rho {\bm u} {\bm u})+{\bnabla}p & = & 
\eta\Delta {\bm u}+\frac{\eta}{3}{\bnabla}({\bnabla}\bcdot{\bm u})+{\bm f},\label{govern}
\end{eqnarray}
where $p=c_{\rm s}^2\rho$ is the pressure, $c_{\rm s}$ is the speed of sound, $\eta>0$ is the dynamic viscosity ($\eta$ is constant 
in space and time in isothermal flows) and $\bm{f}(\bm{x},t)$ is a
random force. Besides the usual conservation of mass and momentum expressed by (\ref{mass}) and (\ref{govern}), 
let us mention two additional ideal integral invariants in unforced isothermal fluids: (i) the total energy density,
\begin{equation}
\langle E\rangle\equiv\langle\rho u^2/2+\rho e\rangle, \label{invare}
\end{equation}
where $e=c_{\rm s}^2\ln(\rho/\rho_0)$ is the specific isothermal compressive potential energy, $\rho_0$ is the mean density of the fluid and
angle brackets $\langle\ldots\rangle$ in this context indicate average over the volume of the fluid, $\frac{1}{V}\int_V(\ldots)\mathrm{d}V$; 
and (ii) the mean kinetic helicity,
\begin{equation}
\langle H\rangle\equiv\left\langle\bm{u}\bcdot\bm\omega\right\rangle\!/2,
\end{equation}
where $\bm\omega\equiv\bnabla\times \bm{u}$ is the vorticity.
In the forced system, the evolution of total energy density is determined by the balance between the action of large-scale force and
small-scale viscous dissipation
\begin{equation}
\partial_t\left\langle E\right\rangle=\left\langle\epsilon\right\rangle-\eta\left\langle\omega^2-
4d^2/3\right\rangle,
\end{equation}
where $\epsilon\equiv\bm{u}\bcdot\bm{f}$ is the local energy injection rate and $d\equiv\bnabla\bcdot\bm{u}$ is the dilatation.

Assuming that a statistical steady state exists at $\Rey\gg1$, the following relation for homogeneous 
turbulence in the inertial interval can be derived \citep{galtier.11}
\begin{equation}
\left\langle d^{\prime}\left(R-2E-p\right) + 
d\,(R^{\prime}-2E^{\prime}-p^{\prime})\right\rangle 
+\bnabla_{\bm{r}}\bcdot\left\langle\left[\delta(\rho\bm{u})\bcdot\delta\bm{u} +
2\delta\rho\delta e\right]\delta\bm{u}
+ \tilde{\delta} e\delta(\rho \bm{u})\right\rangle=-4\varepsilon,\label{key}
\end{equation}
note a different pressure-dilatation term placement \citep{banerjee.13}.
Here $\delta\bm{q}(r)\equiv\bm{q}(\bm{x}^{\prime})-\bm{q}(\bm{x})$ is the increment
in quantity $\bm{q}$ corresponding to the increment $\bm{r}=\bm{x}^{\prime}-\bm{x}$, $r\equiv|\bm{r}|$, 
$R\equiv\rho\bm{u}\bcdot\bm{u}^{\prime}+2\rho e^{\prime}$,
$R^{\prime}\equiv\rho^{\prime}\bm{u}^{\prime}\bcdot\bm{u}+2\rho^{\prime} e$,
$\tilde{\delta}q\equiv q^{\prime}+q$, 
$\bnabla_{\bm{r}}\equiv\hat{\bf e}_i\partial/\partial r_i$
denotes partial derivatives with respect to the increment $\bm{r}$, $\hat{\bf{e}}=\{\hat{\bf e}_1,\hat{\bf e}_2,\hat{\bf e}_3\}$ is an orthonormal basis, 
\begin{equation}
\varepsilon=\langle\bm{u}^{\prime}\bcdot\bm{f} + \bm{u}\bcdot\bm{f}^{\prime}
+ \bm{u}^{\prime}\bcdot\bm{f}\rho^{\prime}/\rho + \bm{u}\bcdot\bm{f}^{\prime}\rho/\rho^{\prime}
\rangle/4
\end{equation}
is the mean energy density injection rate and $\langle\ldots\rangle$ denote an ensemble-average.

Equation (\ref{key}) can be written in symbolic form as
\begin{equation}
S(r)+\bnabla_{\bm{r}}\bcdot\bm{F}=-4\varepsilon, \label{sum}
\end{equation}
where $\bm{F}(r)$ is the total energy flux vector and $S(r)$ represents `source' terms on the left-hand side  
of (\ref{key}) that depend on the potential component of the velocity and can be expressed via increments
\begin{equation}
S(r)=\left\langle\left[\delta(d\rho\bm u)-\tilde{\delta}d\delta(\rho\bm u)\right]\bcdot\delta\bm u+2\left[\delta(d\rho)-\tilde\delta d\delta\rho\right]\delta e + \delta d\delta p-2d p \right\rangle.\label{sour}
\end{equation}
On the basis of empirical evidence (see \S~3 below), in supersonic turbulence the pressure-dilatation 
contribution to the source at $r=0$ is positive, $S(0)=-2\langle d p\rangle>0$.
Since both $e$ and $p$ are proportional to $c_{\rm s}^2$, only the first term in (\ref{sour})
will contribute to $S(r)$ at high Mach numbers ($c_{\rm s}\rightarrow0$)
\begin{equation}
S(r)\approx{\cal S}(r)\equiv\left\langle[\delta(d\rho\bm u)-\tilde{\delta}d\,\delta(\rho\bm u)]\bcdot\delta\bm u\right\rangle\label{sour1}
\end{equation}
\citep[cf. equation (16) in][]{galtier.11}.
Vector relation (\ref{sum}) should be compared to a primitive form of Kolmogorov's (1941) exact and nontrivial four-fifths law,
\begin{equation}
\rho_0\bnabla_{\bm{r}}\bcdot\langle(\delta\bm{u})^2\delta\bm{u}\rangle=-4\varepsilon\label{primitive}
\end{equation}
\citep[e.g.][equation (6.8)]{frisch95}, which follows from (\ref{key}), assuming incompressibility.

If the force in (\ref{govern}) is expressed in terms of the external acceleration, $\bm{f}\equiv\rho\bm{a}$, then
\begin{equation}
\varepsilon=\langle\rho\bm{u}^{\prime}\bcdot\bm{a} + \rho^{\prime}\bm{u}\bcdot\bm{a}^{\prime}
+ \rho^{\prime}\bm{u}^{\prime}\bcdot\bm{a} + \rho\bm{u}\bcdot\bm{a}^{\prime}
\rangle/4.
\end{equation}
In isotropic turbulence,
\begin{equation}
\varepsilon=\langle\rho\bm{u}^{\prime}\bcdot\bm{a} + \rho^{\prime}\bm{u}^{\prime}\bcdot\bm{a} \rangle/2=
\langle\tilde{\delta}\rho\,(\bm{u}^{\prime}\bcdot\bm{a})\rangle/2.\label{eps}
\end{equation}
For incompressible fluids, it can be shown that 
$\varepsilon(r)=\rho_0\langle\bm{u}^{\prime}\bcdot\bm{a}\rangle\approx\rho_0\langle\bm{u}\bcdot\bm{a}\rangle\equiv\rho_0\bar{\varepsilon}$,
if the acceleration $\bm{a}(\bm{x},t)$ (and hence the force $\rho_0\bm{a}$) operate at large scales only. Here, $\bar\varepsilon$ denotes
the (constant) average energy injection rate per unit mass. 

This conventional technique, however, cannot be carried over to compressible fluid turbulence.
In the limit of large correlation length $L_a$ of the acceleration $\bm{a}$, for $r\ll L_a$ the second term in (\ref{eps}) can be 
reduced to a constant $\langle \rho^{\prime}\bm{u}^{\prime}\bcdot\bm{a} \rangle\approx\langle\rho\bm{u}\bcdot\bm{a} \rangle$, while
the first cannot.
If the force $\bm{f}(\bm{x},t)$ had a large correlation length $L_f$ instead, the first term in (\ref{eps}) would reduce to the same combination
$\langle \rho\bm{u}^{\prime}\bcdot\bm{a} \rangle=\langle\bm{u}^{\prime}\bcdot\bm{f}\rangle\approx
\langle\bm{u}\bcdot\bm{f}\rangle=\langle\rho\bm{u}\bcdot\bm{a} \rangle$ for $r\ll L_f$, but the second could not be decoupled simultaneously. 
On the basis of (\ref{eps}), in supersonic turbulence the large-scale external force assumption is incompatible with the presence of inertial interval due to strong density variations on all scales \citep[see also][]{wagner...12}. 
If the large-scale acceleration is also short-correlated in time, then the density can be 
decoupled for $r$ in the inertial interval
$\langle \rho\bm{u}^{\prime}\bcdot\bm{a} \rangle\approx\langle\rho\rangle\langle\bm{u}\bcdot\bm{a}\rangle$ 
\citep[e.g.][]{wagner...12}. In this case, (\ref{eps}) would reduce to 
\begin{equation}
\varepsilon\approx\rho_0\langle\bm{u}\bcdot\bm{a}\rangle=\rho_0\bar{\varepsilon}=\rm const. \label{decouple1}
\end{equation}

In reality, the large-scale acceleration (e.g. the free-fall acceleration in turbulent convection) 
is often not short-correlated in time. Also in simulations, forcing is routinely employed to mimic the energy cascade incoming 
from scales larger than the box size; thus the correlation time cannot be shorter than the large eddy turnover time and it is hard to
expect the decoupling in the form of (\ref{decouple1}).

Nevertheless, the density field in supersonic turbulence has a very short correlation length 
$L_{\rho}\ll L_u\ll L_a$.
Since $\bm{u}$ and $\bm{a}$ are
larger-scale fields, while $\rho$ related to the velocity gradient is a small-scale quantity controlled 
by the nonlinearity of governing equations (\ref{mass}) and (\ref{govern}), one can expect that
$\langle \rho\bm{u}^{\prime}\bcdot\bm{a} \rangle\approx\langle\rho^{\prime}\bm{u}^{\prime}\bcdot\bm{a} \rangle
\approx\langle\rho\bm{u}\bcdot\bm{a} \rangle$. 
Hence, even for $\bm{a}$ with a finite correlation time, the approximation
\begin{equation}
\varepsilon(r)\approx\langle\rho\bm{u}\bcdot\bm{a}\rangle=\varepsilon_0 \label{decouple2}
\end{equation}
is justified for $L_{\rho}\lesssim r\ll L_a$. Using a different approach based on coarse-graining, 
\citet{aluie13} rigorously proved that the energy injection rate is constant at scales sufficiently 
separated from the injection scale, if the external acceleration used to support a statistical 
steady state is restricted to large scales. We will show how well (\ref{decouple1}) and 
(\ref{decouple2}) hold in \S~\ref{exper}.

For now, however, let us assume $\varepsilon(r)=\varepsilon_0$ and integrate (\ref{sum}) over a ball of radius $r$, to obtain an approximate 
scalar relation for isotropic turbulence  in symbolic form:
\begin{equation}
Q(r)+F_{\parallel}(r) \simeq - \frac{4}{3}\varepsilon_0 r,
\label{symbolic}
\end{equation}
where the source function
\begin{equation}
Q(r)\equiv\frac{1}{r^2}\int_0^r S(r)r^2\mathrm{d}r
\label{source}
\end{equation}
and the longitudinal flux of total energy
\begin{equation}
{F}_{\parallel}(r)\equiv\bm{F}\bcdot\bm{r}/r=\left\langle\left[\delta(\rho\bm{u})\bcdot\delta\bm{u} +
2\delta\rho\delta e\right]\delta u_{\parallel}
+ \tilde{\delta}e\delta(\rho u_{\parallel})\right\rangle.
\label{flux}
\end{equation}
The inertial part of the flux dominates at high Mach numbers ($c_{\rm s}\rightarrow0$):
\begin{equation}
F_{\parallel}(r)\approx{\cal F}_{\parallel}(r)\equiv\left\langle\delta(\rho\bm{u})\bcdot\delta\bm{u}\,\delta u_{\parallel}\right\rangle
\label{flux1}
\end{equation}
-- see also (\ref{sour1}). While approximation (\ref{flux1}) does not include the flux of compressive 
energy density $\rho e$, compressibility is still partly accounted for by the momentum difference $\delta(\rho\bm u)$.
Also note that $F_{\parallel}(0)=0$, ${\cal F}_{\parallel}(0)=0$ and $Q(0)=0$.
 
\begin{figure}
\centering
\input{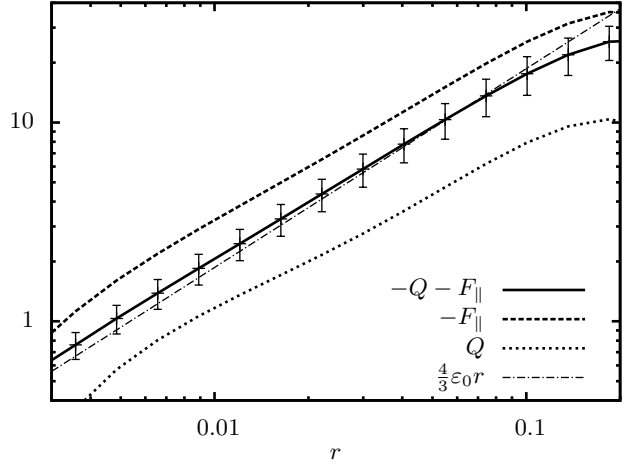}
\vspace{-0.3cm}
\caption{Time-average scaling for the left-hand side and right-hand side of (\ref{symbolic}), including individual contributions from the flux
and source terms. Note that  $F_{\parallel}(r)<0$ (direct energy cascade) and $S(r)>0$ (effective additional energy injection).
Error bars indicate $\pm1\sigma$ variation of the left-hand side of (\ref{symbolic}) in our sample of 86 flow snapshots.}
\label{money}
\end{figure}

\begin{figure}
\centering
\input{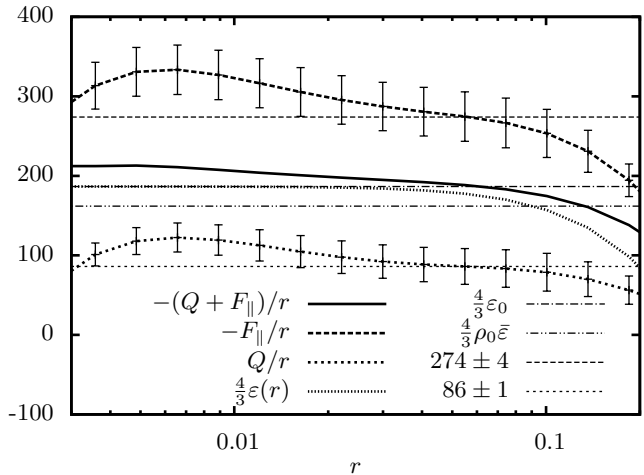}
\vspace{-0.3cm}
\caption{Compensated scaling for the flux and source terms from (\ref{symbolic}). Also shown are various proxies for
the energy injection rate, including the two-point momentum--acceleration and velocity--force correlation functions, as well as
single-point averages with and without density-weighting. Error bars indicate $\pm1\sigma$ variation of the flux and source
terms in our sample.}
\label{comp}
\end{figure}

\section{Numerical verification \label{exper}}
To evaluate (\ref{symbolic}), we shall use data from a numerical experiment designed to study
the inertial range statistics of supersonic homogeneous isotropic turbulence~\citep{K07a}.
The simulation was carried out following the traditional implicit large eddy simulation (ILES) approach 
to the modeling of turbulent flows with strong shocks \citep{grinstein..07}, using an implementation of the 
piecewise parabolic method~\citep[PPM,][]{colella.84} in the Enzo code~\citep{oshea......04}. The forced Euler equations
for an isothermal fluid with the mean density $\rho_0=1$ were numerically integrated in a cubic periodic domain with a linear size 
$L = 1$ covered by a uniform Cartesian grid of $1024^3$ cells. The turbulent r.m.s. Mach number, 
$M \sim6$, and a steady rate of the kinetic energy injection were maintained by a random large-scale
acceleration field, $\bm{a}(\bm{x},t)=C(t)\bm{a}_0(\bm{x})$, with power limited to wavenumbers 
$k/k_{\rm min} \in[1, 2]$, where $k_{\rm min} = 2\pi/L$. The spatially fixed external acceleration $\bm{a}_0(\bm{x})$ was
normalized at every time step to keep the energy injection rate approximately constant in time, $\varepsilon_0=140$; the normalization 
factor $C(t)$ had a standard deviation of $\sim 5\%$ during the simulation. For this work, we used a subset of 
$86$ full data snapshots evenly distributed in the range $t/\tau \in[6,10]$, where the flow crossing
time  $\tau\equiv L/(2c_{\rm s}M)\simeq0.08$. For each snapshot, we computed $\bar\varepsilon$
and evaluated $S(r)$, $Q(r)$, $F_{\parallel}(r)$, and $\varepsilon(r)=\langle\tilde{\delta}\rho\,(\bm{u}^{\prime}\bcdot\bm{a})\rangle/2$ 
for 16 discrete values of increment $r$ from the interval $r/\Delta \in[8, 128]$, where
$\Delta$ is the grid spacing, using $2^{31} \approx 2\times10^9$ randomly selected point pairs for each value of $r$.

Figure~\ref{money} compares the scaling of $-(Q+F_{\parallel})$ with the analytical prediction $4\varepsilon_0 r/3$, 
indicating that the approximate relation (\ref{symbolic}) holds reasonably well. 
Also shown are individual contributions for $Q$ and $F_{\parallel}$. 
As expected for direct energy cascade, the flux is negative across the 
inertial interval. The source function is positive and a factor of $\approx3.2$ smaller than the flux. It represents the 
net effect of mean dilatation at scale $r$ (conditioned on the energy density at this scale) on the associated energy 
flux. The source can be understood as a (positive) correction to $\varepsilon_0$, associated with the evolving metric in
parts of the volume that are subject to compression. In a different context, such `adiabatic heating' of compressible
turbulent fluids was recently considered by \citet{robertson.12}.

Figure~\ref{comp} provides more detail by showing compensated scaling of various terms in (\ref{symbolic});
note that the inertial sub-range is limited to $r\in[0.03,0.1]$ \citep{K07a}. On smaller scales, a bump indicating possible
bottleneck contamination is clearly visible; on larger scales, the action of force is felt directly and 
$\varepsilon(r)$ starts to decline.
The energy injection rate $\varepsilon_0$ defined in (\ref{decouple2}) 
gives an accurate measure for $\varepsilon(r)\approx140$ in the inertial interval, while
(\ref{decouple1}), which ignores the density--velocity correlation, underestimates the injection rate by $\sim15$\%. The
inertial range levels of the flux and source terms, $-274\pm4$ and $86\pm1$ respectively, are estimated from the 
least-squares fits. The source terms play a relatively minor role across the inertial range; both individual flux 
and source contributions in (\ref{symbolic}) scale roughly linearly with $r$, indicating that $S(r)\approx\rm const.$ 
This in turn implies that the kinetic energy cascades conservatively without substantial scale-dependent 
`leakage' to the compressive potential energy \citep[see also][]{aluie..12}.

\begin{figure}
\centering
\input{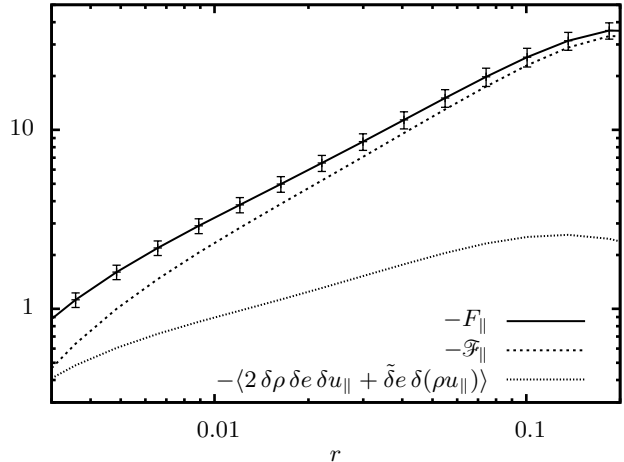}
\vspace{-0.35cm}
\caption{Scaling of the longitudinal flux of total, kinetic and compressive energy. At $M\simeq6$, the kinetic energy flux 
strongly dominates over compressive flux terms. The exponents of the least-squares fits for $r\in[0.03,0.08]$ are 
$0.91\pm0.01$, $0.99\pm0.01$ and $0.44\pm0.02$ for the total, kinetic and compressive fluxes, respectively.
Error bars indicate $\pm1\sigma$ variation of the total flux.}
\vspace{-0.3cm}
\label{fluxes}
\end{figure}

Figure~\ref{fluxes} presents an analysis of different components of the flux in (\ref{flux}). A contribution
from the inertial term ${\cal F}_{\parallel}$ dominates strongly 
on all scales, as predicted by (\ref{flux1}). Two other terms related to the compressive energy flux are subdominant 
in the inertial range. The last term in (\ref{flux}) is $\sim1.6$~dex smaller than the second.

\begin{figure}
\centering
\input{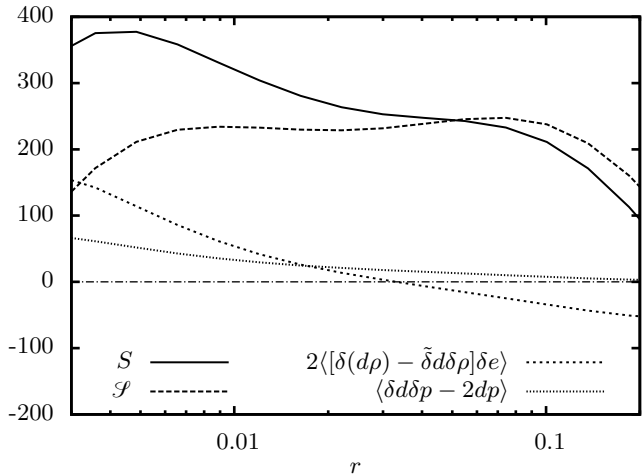}
\vspace{-0.3cm}
\caption{Time-average scaling of various source terms in (\ref{sour}) and (\ref{sour1}).
}
\label{sourfig}
\end{figure}

Figure~\ref{sourfig} shows various constituents of $S(r)$ listed in (\ref{sour}). As suggested by (\ref{sour1}), 
the dominant contribution comes from dynamic-pressure-dilatation terms proportional to $\delta\bm u$; it is positive in a wide range of $r$ and
approximately constant ($\pm4$\%) at $r\in[0.006,0.1]$. Source terms associated with the specific compressive energy difference 
$\delta e$ contribute positively on small scales $r\lesssim0.03$ and act as a sink at $r\gtrsim0.03$. While their effect 
on the inertial range is minimal, they are responsible for an $\approx 40$\% excess in $S(r)$ 
centered around $r\simeq0.004$. The resulting small-scale excess in positive $S(r)$ is in turn responsible for the 
excess in negative $F_{\parallel}(r)$ (see figure~\ref{comp}). The impact of pressure-dilatation terms 
$\langle\delta d\delta p-2dp\rangle$ on $S(r)$ is minor and they can be ignored in the inertial range.
As $r\rightarrow0$, the source is finite and positive: $S(r)\rightarrow -2\langle d p\rangle\simeq73$. 
The average pressure dilatation may depend on various factors such as the Mach number, adopted 
equation of state, numerical resolution and so on.
While for isothermal turbulence at $M\sim6$ we obtain $\langle dp\rangle/(\gamma M^2)\approx-1$, also negative but 
substantially smaller ($\lesssim0.04$) absolute values are reported by \citet{aluie..12} and \cite{wang.....12} for ideal 
gas models at $M\lesssim 1$. At high Mach numbers, the p.d.f. of dilatation is strongly skewed 
towards negative values, making the average pressure dilatation finite and negative.

Since approximations (\ref{sour1}) and (\ref{flux1}) hold at $M\gg1$, (\ref{symbolic}) can be reduced to
\begin{equation}
{\cal Q}(r)+{\cal F}_{\parallel}(r) \simeq - \frac{4}{3}{\cal C}\varepsilon_0 r,
\label{symbols}
\end{equation}
where the source function ${\cal Q}(r)\equiv r^{-2}\int_0^r{\cal S}(r)r^2\mathrm{d}r$ is analogous to (\ref{source}) and
${\cal C}<1$ is a constant of order unity accounting for the fraction of injected kinetic energy that goes into excitation of compressive modes
that were ignored in (\ref{sour1}) and (\ref{flux1}). 
Figure~\ref{super} illustrates the quality of approximations at $M\simeq6$, where ${\cal C}\simeq0.84$, 
i.e. $\approx16$\% of the energy input supports modes related to the compressive energy and pressure dilatation. 
Note that at $M\simeq6$ the right-hand side of (\ref{symbols}) can be replaced by the incompressible expression 
$-4\rho_0\bar{\varepsilon}/3$ without substantial loss of accuracy: see (\ref{decouple1}) for the definition of $\bar\varepsilon$.
As ${\cal S}(r)\approx{\cal S}_{\,0}$ is nearly constant in the inertial interval, 
${\cal F}_{\parallel}(r) \simeq -4\varepsilon_{\rm eff}r/3$,
where $\varepsilon_{\rm eff}={\cal C}\varepsilon_0+{\cal S}_{\,0}/4$.

\section{Discussion}
In practical astrophysical applications, where order-of-magnitude estimates are considered sufficient, relation (\ref{symbols}) can be 
more convenient than (\ref{symbolic}). We therefore explored the scaling of several proxies to ${\cal F}_{\parallel}$, which are 
more closely related to observables and to structure functions previously measured numerically. 
Figure~\ref{proxy} illustrates the scaling of ${\cal F}_{\parallel}(r)$ as well as that of the
transverse $\langle|\delta v_{\perp}|^3\rangle$ and longitudinal $\langle|\delta v_{\parallel} |^3\rangle$ structure
functions of the mass-weighted velocity and compares these with theoretical expectations based on (\ref{symbolic}) and 
(\ref{symbols}). The inertial range slopes of these proxies are close to linear, so they provide a convenient way of estimating 
the value of the total flux $F_{\parallel}$ and thereby the kinetic energy injection rate $\varepsilon_0$. 

Figure~\ref{proxy} shows that  ${\cal F}_{\parallel}(r)\sim 1.7\varepsilon_0$. The total energy flux 
$F_{\parallel}$ would have an $18$\% larger offset, corresponding to a factor of $2$. The proxies based 
on the longitudinal and transverse structure functions of $\bm v$ overestimate $\varepsilon_0$ by factors  $3.5$ and 
$4.1$, respectively. These values can be used to estimate the energy injection rate required to sustain 
turbulent cascade in the interstellar medium of the Milky Way \citep{hennebelle.12}.

\begin{figure}
\centering
\input{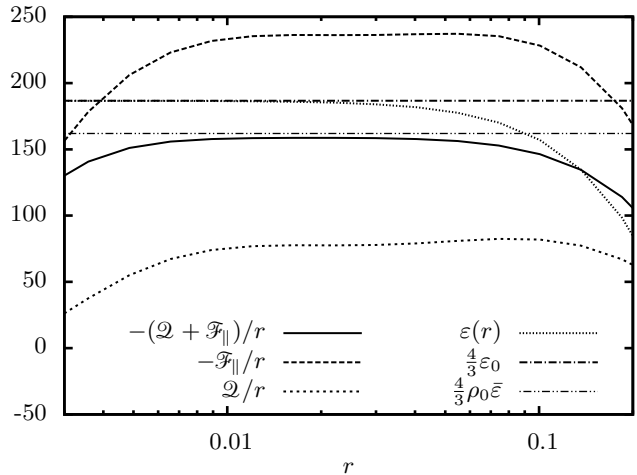}
\vspace{-0.3cm}
\caption{As figure~\ref{comp}, but for the flux and source terms from an approximate relation (\ref{symbols}).
}
\label{super}
\end{figure}

\begin{figure}
\centering
\input{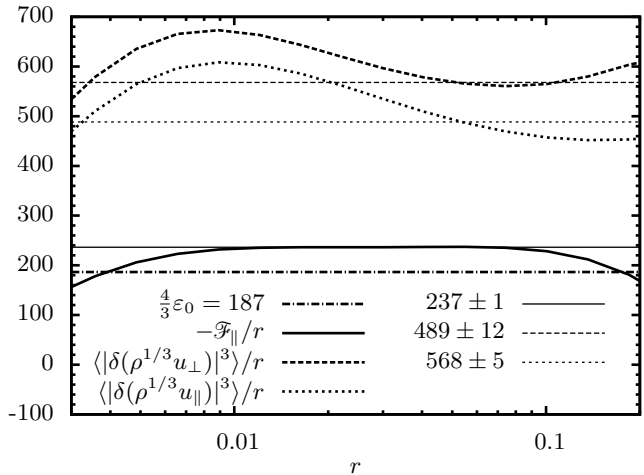}
\vspace{-0.3cm}
\caption{Compensated scaling for various proxies of the energy injection rate $\varepsilon_0$.
Thin lines show horizontal least-squares fits at $r\in[0.03,0.1]$; numbers indicate the best-fit value and its
standard deviation.}
\label{proxy}
\end{figure}

An interesting feature to notice in figures~\ref{super} and \ref{proxy} is the extended ($\sim1$~dex) 
linear scaling range of ${\cal F}_{\parallel}$, 
which continues down to $r\simeq0.008\simeq8\Delta$. In ILES carried out with a PPM-based
code, all scales below $\sim16\Delta$ are strongly affected by numerical dissipation \citep{porter.94}. 
Scales shorter than $\sim32\Delta$ are usually identified with the so-called `bottleneck bump' \citep{falkovich94,porter..94}, where the 
energy piles up in the near-dissipation part of the inertial range due to a steep wavenumber dependence of numerical diffusivity in the 
dissipation range \citep[$\propto k^{4-5}$ for PPM, see][]{porter..92}. 
Similar bumps are present in the spectra of velocity, density and various mixed quantities 
in supersonic turbulence \citep{K07a}. In structure functions, the bottleneck is expected to be more pronounced at higher orders 
\citep{falkovich94}; it is also less localized than in power spectra due to mixing of small- and large-scale information 
\citep[see, e.g., a plot of $\langle|\delta v_{\perp} |^3\rangle$ in figure~\ref{proxy} and note that we take an absolute
value]%
{dobler...03,davidson.05}. The inertial flux ${\cal F}_{\parallel}$, however, does not show a bump, as expected in the
inertial cascade, when an absolute value operation is not applied. 

As we discussed above, the bump in the total flux is associated with the compressive energy flux contribution
$\langle2\delta\rho\delta e\delta u_{\parallel}\rangle$, which becomes comparable to the kinetic energy flux and 
also somewhat flattens below the sonic scale at $r\lesssim r_{\rm s}$: $\delta u_{\parallel}(r_{\rm s})=c_{\rm s}$. 
This behaviour may depend on the details of the shock-capturing scheme. Indeed, a PPM 
implementation in the Enzo code \citep{oshea......04} produces a growing fraction of dilatational modes in the 
velocity power spectrum on scales below $32\Delta$ \citep[see figure 1b in][]{kritsuk...10}, which could 
potentially contribute to the bump build-up. 

\section{Conclusions and final remarks}
We verified a relation for correlation functions in compressible isothermal turbulence \citep{galtier.11}
with data from a numerical simulation at Mach 6 \citep{K07a}. While an isotropic version of the relation is not strictly exact, it provides a good approximation to numerical results. Our analysis of different terms in (\ref{symbolic}) 
supports a Kolmogorov-like picture of the energy cascade in supersonic turbulence previously discussed on a phenomenological level \citep{K07a} and recently supported theoretically \citep{aluie11,aluie..12,aluie13}. 
A non-trivial new approximate relation (\ref{symbols}) that holds at high turbulent Mach 
numbers is proposed. The relation represents an important step beyond phenomenology, as it 
sheds light on the problem of universality in compressible turbulence and provides a way to quantitatively 
predict the energy injection rate from the scaling of certain combinations of observables. This result can have 
important implications for interstellar turbulence, as approximately constant energy transfer rates are observed 
in the ISM over more than four decades in length scale \citep{hennebelle.12}. 

The fourth-order scaling relation (\ref{symbolic}) traditionally formulated in terms of the energy flux is not the only compressible analogue of Kolmogorov's four-fifths law. Another approximate relation for homogeneous isothermal turbulence, formulated in terms of fluxes and densities of conserved quantities,
\begin{equation}
\bnabla_{\bm{r}}\bcdot\left\langle\rho\rho^{\prime}\left[\left(\bm{u}\bcdot\bm{u}^{\prime}\right)\bm{u}^{\prime}+c^2_{\rm s}\bm{u}\right]\right\rangle
\sim-\left\langle\rho\rho^{\prime}\right\rangle\bar{\varepsilon},\label{wagner}
\end{equation}
cf. (\ref{key}), has recently been obtained and verified with the same numerical data \citep{wagner...12}.
More strictly, this fifth-order flux relation should be viewed as an anisotropic analogue of the von K\'arm\'an--Howarth 
relation, as it involves correlation functions, but (\ref{wagner}) can also be reduced to the four-fifths law in the 
incompressible limit \citep{falkovich..10}. 
Note that the dependence of the density autocorrelation function in the right-hand side of (\ref{wagner}) 
on the increment $r$ varies with the Mach number, as does the slope of the density power spectrum \citep{kim.05}.
Unlike (\ref{symbolic}), an isotropic version of (\ref{wagner}) does not have a trivial right-hand side universally linear in $r$. 
In a particular case at $M\simeq6$, the density autocorrelation function has a logarithmic dependence on $r$ and 
a closed-form analytical representation of the isotropic flux relation is feasible \citep[see (3.3) in][]{wagner...12}. 

We thus conclude that at least two compressible analogues of Kolmogorov's four-fifths law exist, consistent with the 
extension of the turbulent energy cascade picture to supersonic regimes. 
Only the fourth-order energy cascade relation (\ref{symbolic}) is `universal' in the sense that its right-hand side remains approximately 
linear in the inertial range at all Mach numbers. It is worth noting that the fourth-order relation exploits the 
conservation of total energy (which is an inviscid invariant), while the fifth-order one follows from conservation 
of momentum and involves the momentum density and flux. 
\\

We thank Hussein Aluie,  Gregory Falkovich and S\'ebastien Galtier for stimulating discussions.
This research is supported in part by NSF grants AST-0908740 and AST-1109570.  
The simulation utilized TeraGrid computer time allocations MCA98N020 and MCA07S014 at SDSC.
The analysis was performed on the XSEDE resource Gordon at SDSC
under a Director's Discretionary Allocation. 

\bibliographystyle{jfm}

\end{document}